          [2001/04/25 1.1 (PWD)]
\documentclass[a4paper,twocolumn]{esapub} 

\usepackage{times}

\usepackage{natbib,amssymb,graphicx,hyperref}

\title{Gamma-ray bursts and X-ray melting of material as a
potential source of chondrules and planets.}

\author{Paul Duggan$^1$,
        Brian McBreen \footnote{Correspondence to \textit{brian.mcbreen@ucd.ie}} $^1$,
        Alun J. Carr$^2$,
        Sheila McBreen$^1$,
        Elaine Winston$^1$,
        Lorraine Hanlon$^1$, and\ 
        Leo\,Metcalfe$^{3}$}
\affil{$^1$Department of Experimental Physics, University College
Dublin, Ireland.\\ $^2$Department of Mechanical Engineering,
University College Dublin, Ireland.\\
$^3$ XMM-Newton Science Operations Centre, European Space Agency,
Villafranca del Castillo, PO Box 50727, 28080 Madrid, Spain.}

\def\degr{\hbox{$^\circ$}}
\newcommand{\up}[1]{\raisebox{1.5ex}[0cm][0cm]{#1}}

\begin{document}


\maketitle

\begin{abstract}
The intense radiation from a gamma-ray burst (GRB) is shown to be
capable of melting stony material at distances up to 300 light
years which subsequently cool to form chondrules. These conditions
were created in the laboratory for the first time when millimeter
sized pellets were placed in a vacuum chamber in the white
synchrotron beam at the European Synchrotron Radiation Facility
(ESRF). The pellets were rapidly heated in the X-ray and gamma-ray
furnace to above 1400\,\degr C melted and cooled. This process
heats from the inside unlike normal furnaces. The melted spherical
samples were examined with a range of techniques and found to have
microstructural properties similar to the chondrules that come
from meteorites. This experiment demonstrates that GRBs can melt
precursor material to form chondrules that may subsequently
influence the formation of planets. This work extends the field of
laboratory astrophysics to include high power synchrotron sources.
\end{abstract}

\begin{figure}[p]\centering
\includegraphics[width=\columnwidth]{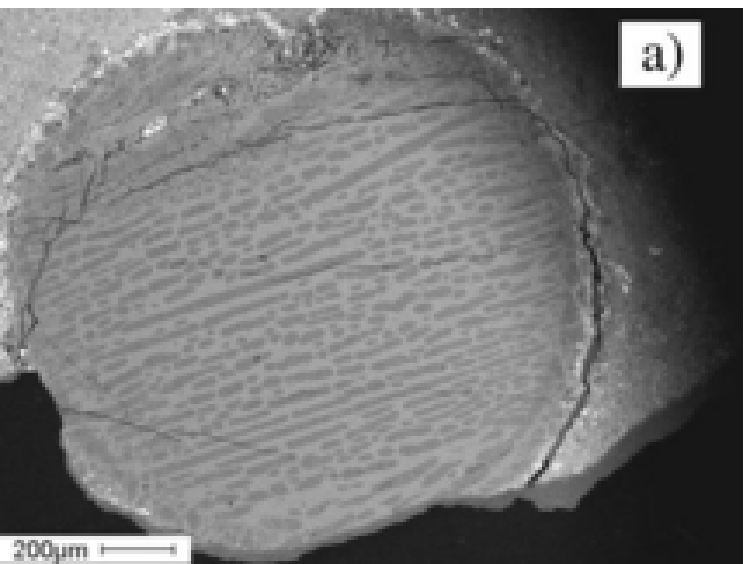}\\[3pt]
\includegraphics[width=\columnwidth]{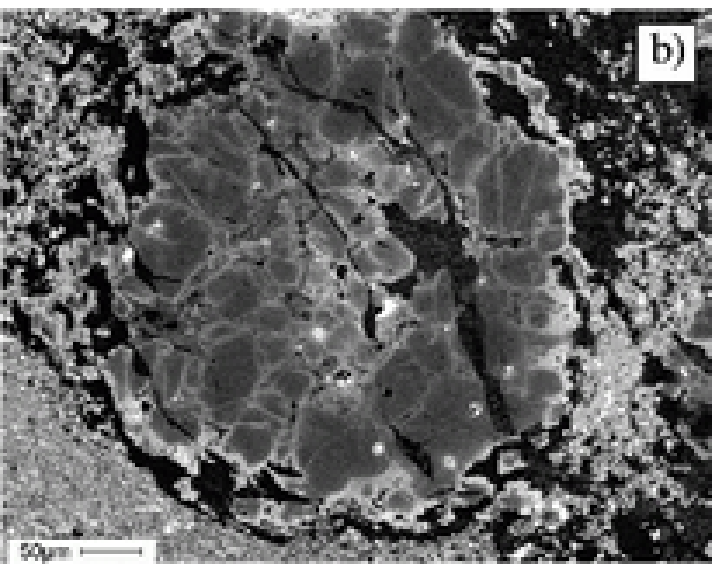}
    \caption{Backscattered electron microscope images of two
chondrules from the Allende meteorite. Phases with higher atomic
number are brighter in color. The dark grey grains are elongated
olivine crystals in (a) and porphyritic crystals in (b) where the
brighter regions are the interstitial glassy material. The
diameters of the chondrules are about 1\,mm and 0.6\,mm and are
surrounded by matrix material in the meteorite.}
    \label{fig:met}
\end{figure}

\begin{figure*}[t]
  \includegraphics[width=0.32\textwidth]{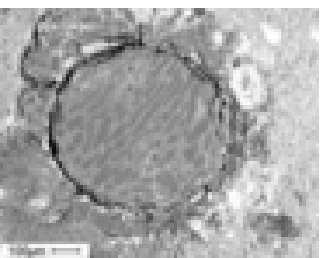}
    \includegraphics[width=0.32\textwidth]{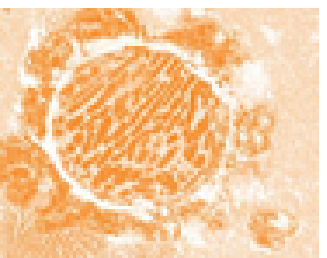}
    \includegraphics[width=0.32\textwidth]{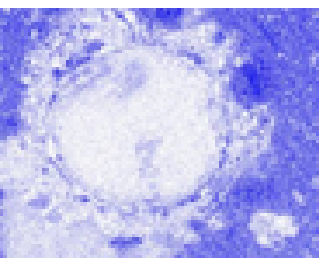}\\[5pt]
    \includegraphics[width=0.32\textwidth]{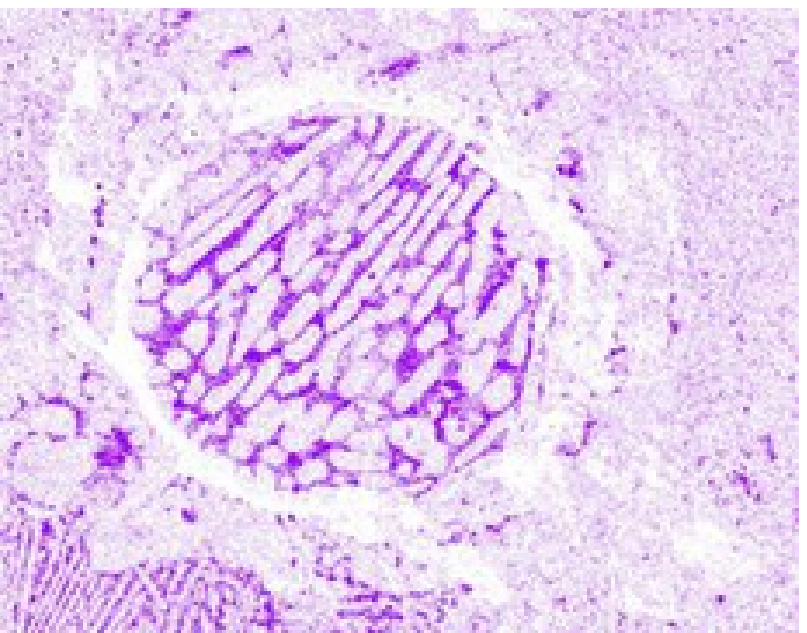}
    \includegraphics[width=0.32\textwidth]{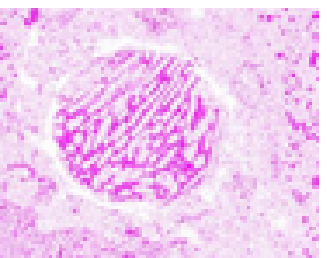}
    \includegraphics[width=0.32\textwidth]{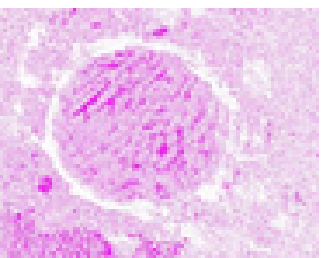}
\caption{Backscattered electron microscope image and X-ray maps of
a chondrule from the Allende meteorite. In the case of the X-ray
elemental maps, regions with greater proportions of an element
have more intense colour. These images are created by analysing an
X-ray window around the K$_\alpha$ spectral line for each element
during the raster scan by the electron beam. Top row from left:
(a) backscattered electron image, (b) Mg X-ray map and (c) Fe
X-ray map. Bottom row from left: (d) Al X-ray map, (e) Ca X-ray
map and (f) Na X-ray map. The Mg is preferentially located in the
olivine crystals (b) and the glassy material between the olivine
crystals is enriched with Al and Ca (d and e). There is more iron
in the matrix material surrounding the chondrule (c) as shown by
the dark region. }\label{fig:metxray}
\end{figure*}

\section{Introduction}
Chondrules are millimetre-sized, stony spherules
(Fig.\,\ref{fig:met}) that constitute the major component of most
chondritic meteorites that originate in the region between Mars
and Jupiter and which fall to the Earth. They appear to have
crystallised rapidly from molten or partially molten drops and
were described \citep{sorby1877} as ``molten drops in a fiery
rain''. The properties of the chondrules and chondrites have been
exquisitely deduced from an extensive series of experiments
\citep{wasson2003,hewins2004} and two conferences have been
devoted completely to chondrules \citep{king1983,hewins1996}. The
mineralogy of chondrules is dominated by olivine
((Mg,Fe)$_2$SiO$_4$) and pyroxene ((Mg,Fe)SiO$_3$) and there is a
wide range of compositions for all elements. A compositional map
of a chondrule is shown in Fig.\,\ref{fig:metxray}. The magnesium
is preferentially found in the dark olivine crystals
(Fig.\,\ref{fig:metxray}(b)) while the aluminium and calcium are
located in the glassy material between the crystals
(Fig.\,\ref{fig:metxray}(d) and Fig.\,\ref{fig:metxray}(e)). The
diversity of chondrules is consistent with the melting of
heterogeneous precursor solids or dust balls. The age of
chondrules indicate they formed very early in the solar system.
The calcium-, aluminium-rich inclusions (CAIs) are refractory
inclusions in carbonaceous and ordinary chondrites that predate
the chondrules by several million years and are the oldest known
solid materials produced in the nebula \citep{swindle1996}. The
first 10$^7$ years in the complicated development of the solar
system has been comprehensively covered in an attempt to
understand the CAI to chondrule time interval of several million
years \citep{cameron1995}.

The presence of volatile elements in the chondrules
(Fig.\,\ref{fig:metxray}(f)) indicate that the high temperature
melting period lasted for a matter of seconds to minutes.
Experiments based on chemical and textural compositions of
chondrules suggest cooling rates that were much slower than
radiative cooling of isolated chondrules and imply they were made
in some large quantity in relatively opaque nebular domains
\citep{yu1998,brearley1998}. Volatile elements such as alkalis and
sulphur occur in chondrule interiors as primary constituents and
indicate that some chondrule precursor materials must have reacted
with cool nebula gases at ambient temperatures less than 650\,K.

The identification of the heat source responsible for chondrule
formation is important for the understanding of the formation of
planets because of the large amount of chondrules present in
chondrite meteorites. The heat source remains uncertain and a
critical summary of the heating mechanisms was given by
\citet{rubin2000}. These methods include bipolar outflows
\citep{sekiya1998,lee1998} and shock wave heating of the precursor
materials \citep{wood1988,miura2004,ciesla2002}. Almost all
proposed heat sources \citep{rubin2000,boss1996,jones2000} are
local to the solar nebula, one exception being the proposal
\citep{mcbreen1999,duggan2001} that the chondrules were flash
heated to melting point by a nearby GRB when the iron all through
the precursor material efficiently absorbed X-rays and low energy
$\gamma$-rays. The distance to the source was about 300 light
years (or 100\,pc) for a GRB output of 10$^{53}$\,erg and was
estimated using the minimum value of
2\,$\times\,10^{10}$\,erg\,g$^{-1}$ required to heat and melt the
precursor grains \citep{grossman1988,wasson1993}. The role of
nearby supernovae that preceded the formation of the solar system
have been considered \citep{cameron1995} along with the serious
consequences for life on Earth of nearby supernovae
\citep{ruderman1974,clark1977} and GRBs
\citep{thorsett1995,scalo2002}. The consequences of a nearby GRB
on the early solar nebula have not been considered elsewhere.

\section{GRBs as a heat source}

\begin{figure}[t]
\centering
    \includegraphics[width=\columnwidth]{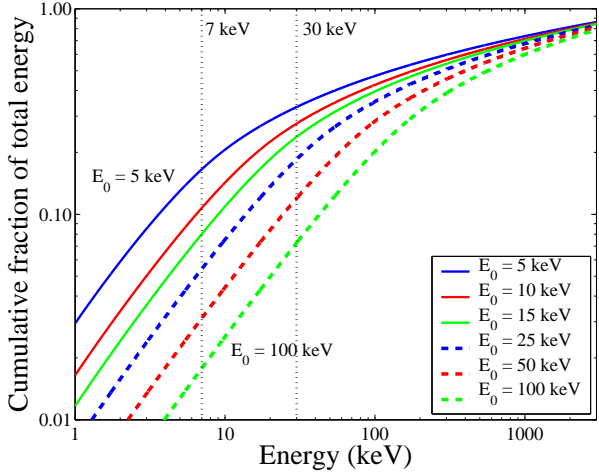}
        \caption[The cumulative fraction of GRB energy as a function of
                photon energy.]{The cumulative fraction of GRB energy as a function of
        photon energy for assumed spectral parameters $\alpha$= -1,
                $\beta$= -2, E$_0$=5, 10, 15, 20, 25, 50 and 100\,keV and
                redshift of 0.8.}
    \label{fig:cumulative_energy}
\end{figure}

Since their discovery thirty years ago the properties of GRBs have
been determined by an outstanding series of experiments that were
deployed on more than twenty spacecraft at distances up to several
astronomical units (AU) from the Earth and during one period 11
spacecraft were used to study the same GRBs. The properties of
GRBs have been reviewed
\citep{fishman1995,piran1999,hurley2002,zhang2003} and are the
subject of intense research. GRBs are extragalactic in origin and
release collossal amounts of energy, 10$^{52}$ to 4\,$\times
10^{54}$\,erg assuming isotropic emission, for the GRBs with known
redshift. 
The index of the X-ray afterglow is in the range t$^{-1.1}$ to t$^{-1.5}$
\citep{costa1997,piro2004,nicastro1998}. Simultaneous optical
emission was detected from the spectacular GRB 990123 at the level
of about 10$^{-5}$ of the energy in $\gamma$-rays
\citep{akerlof1999}, but this emission is too weak to influence
chondrule formation.

\begin{figure}[t]
\centering
    \includegraphics[width=\columnwidth]{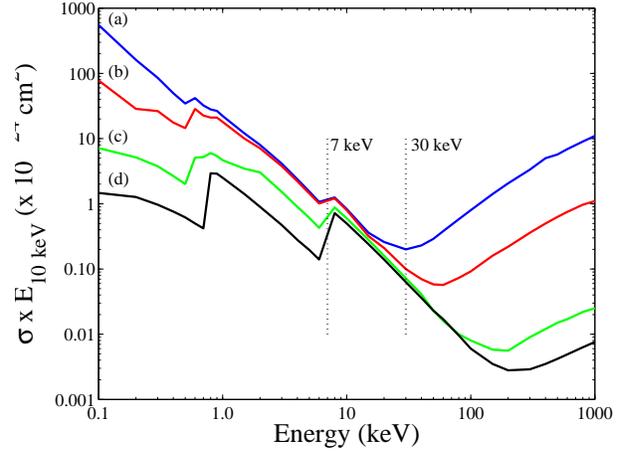}
    \caption[The combined photoelectric and Compton cross-sections
        relative to H as a function of energy]{The combined photoelectric
    and Compton cross-sections relative to H as a function of energy and
    scaled by E/10\,keV for clarity of presentation.  (a) The elements H
    to Fe that are at least as abundant as Fe (H, He, C, N, O, Ne, Mg,
    Si, Fe) (b) the same elements as in (a) with solar abundance of H
    and He reduced by 10\% (c) The precursor dust combination
    Mg$_{1.1}$Fe$_{0.9}$SiO$_{4}$ and (d) the element Fe.}
\label{fig:abs_cross_sect}
\end{figure}

The photon spectra of GRBs are well described by a power-law with
a low energy slope $\alpha$, a break energy E$_0$ and a high
energy power law with slope $\beta$. The functional form
is given by \citep{band1993}:

$N(E)\,=\,A E^\alpha e^{-E/E_0}$ and $N(E)\,=\,BE^\beta$,
$\alpha\,>\,\beta$.

The value of E$_0$ ranges from 2\,keV to over 1\,MeV and the
indices $\alpha$ and $\beta$ are typically -1 and -2 respectively.
It is interesting to note that the temporal properties of the long
and short GRBs are more complex than their spectral properties
 and are well described by a lognormal
distribution \citep{mcbreen2001,quilligan2002,smcbreen2002}.

The spectral energy distributions of a sample of GRBs have been
extrapolated and integrated from 0.1\,keV to 10\,MeV using a
sample of values for E$_0$ compatible with BATSE \citep{band1993}
and Ginga results \citep{strohmayer1998}. The cumulative fraction
of the total energy in GRBs is given in
Fig.\,\ref{fig:cumulative_energy} where a redshift correction of
z\,=\,0.8 has been applied to all the spectra.

\begin{figure}[t]
\centering
    \includegraphics[width=\columnwidth]{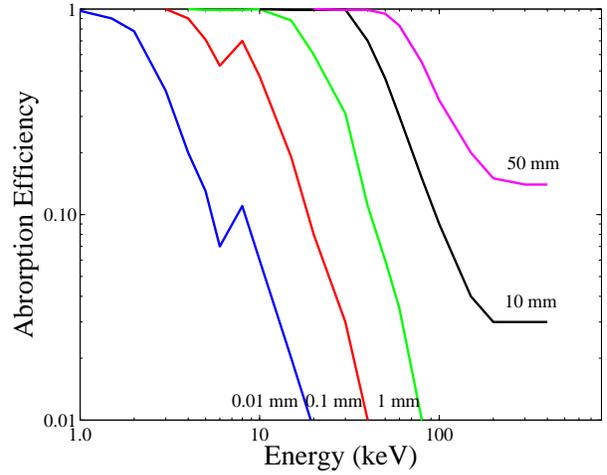}
    \caption[The absorption efficiency of different thickness of
    $Fe_{0.9}Mg_{1.1}SiO_4$\ as a function of energy] {The
    absorption efficiency of different thickness of
    $Fe_{0.9}Mg_{1.1}SiO_4$\ as a function of energy: (a) 0.01\,mm,
    (b) 0.1\,mm, (c) 1\,mm, (d) 10\,mm, (e) 50\,mm.}
    \label{fig:abs_eff}
\end{figure}

\section{Attenuation of X-rays and $\gamma$-rays}

The absorption of the GRB energy by the gas and dust in the nebula
would have occurred through the processes of photoelectric
absorption and Compton scattering. The combined cross-sections due
to these processes for the elements from H to Fe are plotted in
Fig.\,\ref{fig:abs_cross_sect}. Solar abundance values were
adopted for the nebula \citep{anders1989} and the photoelectric
and Compton cross-sections for the elements have been used
\citep{veigele1973}. The photoelectric effect absorbs the photon
completely but in Compton scattering only a fraction of the energy
is removed per scattering and this fraction varies from 0.14 at
100\,keV to 0.45 at 1\,MeV. The product of the Compton
cross-section by the fraction of the energy absorbed in the first
scattering was used for the Compton cross-section and many
scatterings may occur before the degraded photon is finally
absorbed by the photoelectric effect. Decreasing the abundance of
H and He by a factor of 10 relative to the solar value
significantly reduces the cross-section below 1\,keV and above
20\,keV (Fig.\,\ref{fig:abs_cross_sect}(b)). The Fe cross-section
(Fig.\,\ref{fig:abs_cross_sect}(d)) is dominant between the K
absorption edge at 7.1\,keV and 30\,keV but the upper bound of
30\,keV extends to above 50\,keV for low abundance of H and He. Fe
makes the major contribution to the absorption by the dust and is
the key to chondrule formation.

\begin{table}[t]
\caption{Elemental oxide compositions of three
types of precursor material in weight percentages and the
calculated liquidus temperatures. The compositions are the same as
listed \citep{yu1998} except that the small amounts of K$_2$O
($<$\,0.11\,\%) were not included and magnetite (Fe$_3$O$_4$) was
used in place of FeO.}\label{tab:comp}
 \centering
\begin{tabular}{lccc} \hline\hline
            & \underline{Type\,IA}& \underline{Type\,IAB} &
\underline{Type\,II} \\
SiO$_2$         &   45.4   & 47.2  &  49.2  \\
TiO$_2$         &   0.1    & 0.1   &  0.1   \\
Al$_2$O$_3$     &   4.9    & 9.7   &  4.1   \\
Fe$_3$O$_4$     &   8.3    & 6.6   &  21.3  \\
MnO             &   0.1    & 0.1   &  0.1   \\
MgO             &   37.4   & 30.7  &  22.7  \\
CaO             &   2.5    & 3.5   &  0.2   \\
Na$_2$O         &   1.3    & 2.1   &  2.3
\\[2pt]
Liquidus   \\
\,\,Temperature     & \up{1,692\,\degr C}     &\up{1,577\,\degr C}
&
\up{1,509\,\degr C}\\
\hline\hline
\end{tabular}
\end{table}

A composition consisting of solar abundance of oxides of Fe, Si
and Mg or Mg$_{1.1}$Fe$_{0.9}$SiO$_4$ was assumed for the
precursor grains and the product of the cross-section of this
combination by abundance relative to H is plotted in
Fig.\,\ref{fig:abs_cross_sect}(c). The X-ray and $\gamma$-ray
absorption efficiencies of different thicknesses of precursor
grains, assuming a density of one, are given in
Fig.\,\ref{fig:abs_eff} and grains in the range 
0.01\,mm to 50\,mm
are very efficient absorbers in the region where dust dominates
the absorption (Fig.\,\ref{fig:abs_cross_sect}). This range agrees
quite well with the measured Weibull and lognormal distributions
of chondrule sizes \citep{martin1980,rubin1984}. The deficiency of
small grains is caused by low absorption efficiency and
substantial radiation losses from grains with large surface to
volume ratios. The thickness of the dust layer converted to
chondrules depends on the GRB spectrum which must have significant
emission below 30\,keV where dust absorption dominates
(Fig.\,\ref{fig:abs_cross_sect}) and also on the mixture and
distribution of gas and dust in the nebula. For a GRB with
10$^{53}$\,erg and an assumed spectrum $\alpha$\,=\,-1,
$\beta$\,=\,-2 and E$_0$\,=\,15\,keV
(Fig.\,\ref{fig:cumulative_energy}), the fraction of GRB energy
photoelectrically absorbed by the dust is 20\%, increasing to 27\%
for a factor 10 reduction in H and He. In the simplified case of
solar abundance and a uniform mix of gas and dust, the thickness
of the chondrule layer created is 0.18\,g\,cm$^{-2}$ corresponding
to one optical depth for 30\,keV X-rays. The layer thickness
increases to about 0.8\,g\,cm$^{-2}$ and 2.0\,g\,cm$^{-2}$ for
optical depths to 40\,keV and 55\,keV X-rays with H and He
abundances reduced by factors of 3 and 10 respectively. The
thickness of the chondrule layer is therefore controlled by the
degree of gas depletion from the nebula. The minimum GRB fluence
required to produce chondrule layers of 0.18, 0.8 and
2.0\,g\,cm$^{-2}$ is 1.8\,$\times\,10^{10}$,
7.0\,$\times\,10^{10}$ and
1.5\,$\times\,10^{11}$\,erg\,cm\,$^{-2}$, adopting 20\%, 23\% and
27\% absorption by the chondrule precursors and
2$\times\,10^{10}$\,erg\,g$^{-1}$ for heating and melting. A
fluence of 10$^{11}$\,erg\,cm$^{-2}$ implies a distance of about
100\,pc to the source for an output of 10$^{53}$\,erg radiated
isotropically. The GRB would also form a layer of chondrules over
a large area (10$^3$ -- 10$^4$\,pc$^2$) in a nearby molecular
cloud provided large precursor grains had already formed
\citep{weidenschilling1994}. The process of chondrule amalgamation
might be sufficient to trigger star formation over this region. In
this case chondrule formation precedes cloud collapse and star
formation. The existence of pre-solar grains in meteorites is well
established \citep{zin1996} but there is no evidence for pre-solar
chondrules.

\begin{figure}[t]\center
\includegraphics[width=\columnwidth]{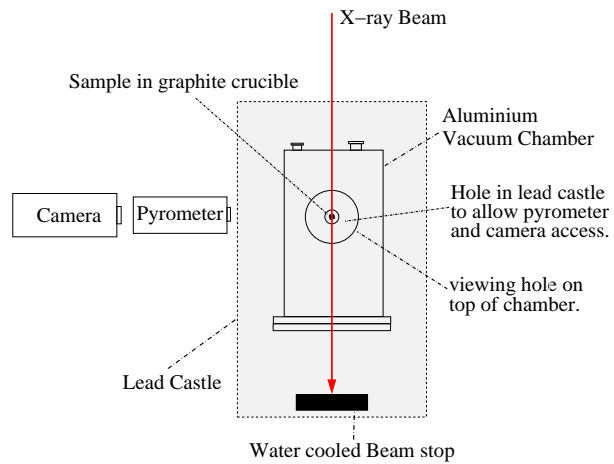}
\caption{Schematic diagram of experimental setup at the ESRF. The
lead castle shield encasing the experiment is shown by the grey
box.}\label{fig:expsetup}
\end{figure}

The distance to the source could be up to 300 light years
($\sim$\,100\,pc) for a GRB with an isotropic output of
10$^{53}$\,erg \citep{piran1999,mes2002}. This distance limit was
obtained using the minimum value of 2\,x\,10$^{10}$\,erg\,g$^{-1}$
required to heat and melt the precursor grains \citep{wasson1993}
and is equivalent to an enormous energy deposition of
10$^{11}$\,erg\,cm$^{-2}$.

\section{Experimental Method}
It is now possible to create the astrophysical conditions near a
GRB source in the laboratory due to the development of powerful
synchrotrons \citep{duggan2003}. The ESRF has a 6\,GeV,
third-generation synchrotron capable of generating the required
power. A wiggler device was inserted and used to create X-rays in
the range 3 -- 200\,keV. The 24-pole wiggler has a characteristic
energy of 29\,keV at a minimum wiggler gap of 20.3\,mm. Time was
awarded on the ID11 white beam to test the prediction that large
fluxes of X-rays and $\gamma$-rays could melt millimetre sized
dust grains and hence extend the use of high power synchrotron
sources to laboratory astrophysics \citep{remington1999}. The
composition of chondrules varies widely and a classification
system based on the iron content is often used
\citep{mcsween1977,yu1998}. Type IAB chondrules have low iron
content while type IA and type II have increasing amounts of iron.
The pellets were made from a mixture of elemental oxides with
weight percentages as given in Table\,\ref{tab:comp} for the three
precursors types.

The major effect  of including more iron is to reduce the
magnesium content and the liquidus temperature of the material.
The oxides, without the volatile Na$_2$O, were mixed and heated to
400\,\degr C in an alumina crucible for 13 hours. The powders
after heating had a mass loss of about 5\,\%, due to moisture loss
and reduction of the elemental oxides. After cooling, Na$_2$O was
added to the mixture. The powder was pressed into cylindrical
pellets of diameter 3\,mm and height of 3\,mm.

\begin{figure}[t]\centering
\includegraphics[width=\columnwidth]{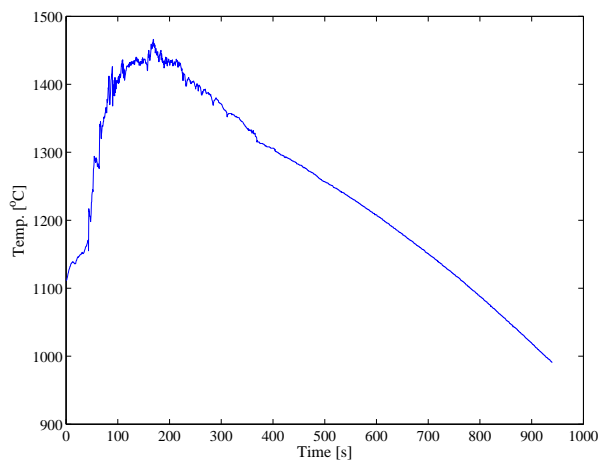}
    \caption{The temperature profile of a typical sample in the
synchrotron experiment. The reason for the high initial
temperature is that 1000\,\degr C is the minimum value read by the
pyrometer.  The temperature rose rapidly to above 1000\,\degr C
when the beam shutter was opened.}
    \label{fig:temp_profile}
\end{figure}

\begin{figure}[p]
\centering
\includegraphics[width=\columnwidth]{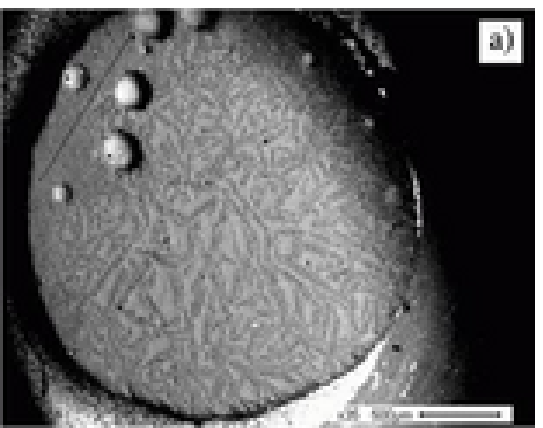}\\[3pt]
\includegraphics[width=\columnwidth]{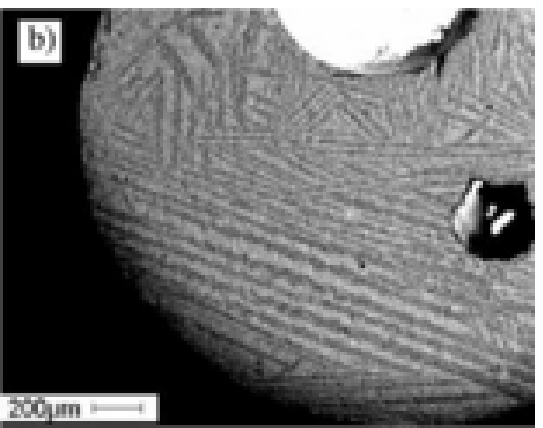}\\[3pt]
\includegraphics[width=\columnwidth]{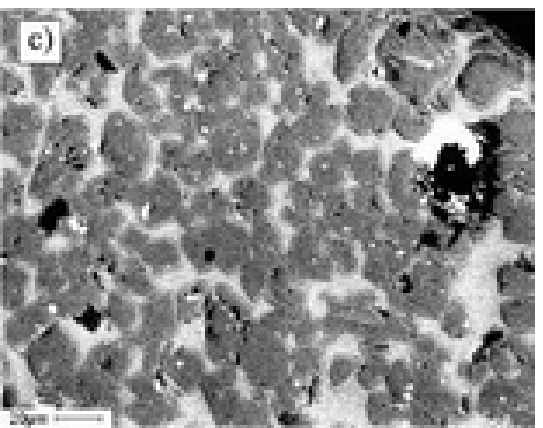}
    \caption{Backscattered electron microscope images of three melted
samples from the synchrotron experiment. (a) cross-section of a
complete sample with randomly orientated olivine crystals and type
IA precursor composition. (b) section of a sample with barred
olivine crystals and type II precursor composition (c) section of
a sample with porphyritic crystals and type IAB precursor
composition. The samples in (a) and (b) have cavities of size $<$
500\,mm. Cavities have been observed \citep{maharaj1994} in other
experimentally produced samples and are caused by trapped gases
and incomplete melting.}
    \label{fig:sync}
\end{figure}

\begin{figure*}[t]
    \includegraphics[width=0.32\textwidth]{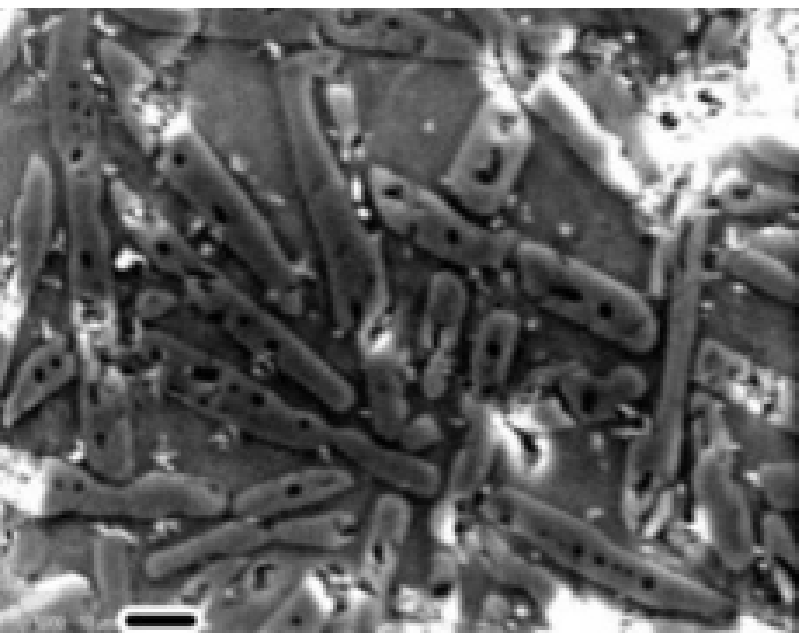}
    \includegraphics[width=0.32\textwidth]{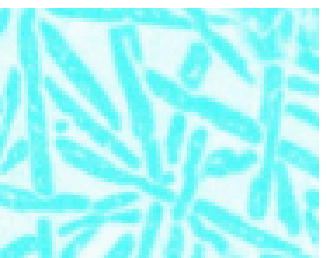}
    \includegraphics[width=0.32\textwidth]{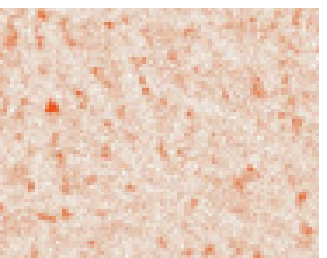}\\[5pt]
    \includegraphics[width=0.32\textwidth]{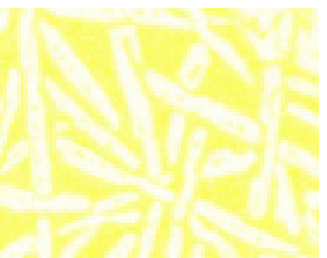}
    \includegraphics[width=0.32\textwidth]{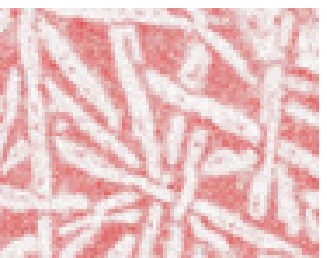}
    \includegraphics[width=0.32\textwidth]{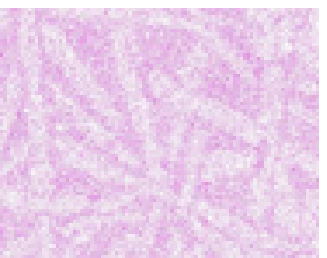}
\caption{Electron microscope image and X-ray maps  of a section of
type IA X-ray heated sample. Top row from left: (a) secondary
electron image with $\times$1000 magnification (the black bar in
the bottom left hand corner measures 10\,$\mu$m), (b) Mg X-ray map
and (c) Fe X-ray map. Bottom row from left: (d) Al X-ray map, (e)
Ca X-ray map and (f) Na X-ray map. The partitioning of the
elements is similar to that shown in the chondrule in
Fig.\,\ref{fig:metxray}. The Mg is concentrated in the olivine
crystals and the Al and Ca are enhanced in the glassey material
between the crystals.}\label{fig:samp_xray}
\end{figure*}

Each pellet was placed in a graphite crucible inside an evacuated
container and inserted into the path of the white X-ray beam
(Fig.\,\ref{fig:expsetup}). The size of the beam was 2\,mm x
1.5\,mm. The synchrotron beam entered the vacuum chamber through a
Kapton window of thickness 0.05\,mm. The pressure in the container
was between 10$^{-2}$ -- 10$^{-3}$\,mbar, which is typical of
planetary forming systems. In a few cases the residual air in the
vacuum chamber was replaced with hydrogen. During the heating
cycle the temperature of the pellet was measured using a Raytex
MR1SCSF pyrometer with a range from 1000\,\degr C to 3000\,\degr
C. The pyrometer was located outside the lead shielding and viewed
the sample via a mirror through a glass window on the top of the
vacuum container and at right angles to the X-ray beam. This
window had to be replaced on several occasions during the
experiment because of darkening caused by radiation damage. The
sample was also monitored with a camera that viewed it through the
pyrometer optics. The pellets were rapidly heated in the X-ray and
$\gamma$-ray furnace to temperatures above 1400\,\degr C
(Fig.\,\ref{fig:temp_profile}). During the heating and melting
process the pellets bubbled, moved about and sometimes ejected
small drops of iron rich material. The melted samples were kept at
the maximum temperature for a duration of 10\,s to 300\,s and
cooled when the power in the beam was reduced by widening the
gap between the magnets of the wiggler. The beam was removed when the temperature
dropped below 1000\,\degr C. The samples cooled rapidly to yield
2\,--\,3\,mm diameter black spherules.

A Huber model 642 Guinier X-ray powder diffractometer with
monochromatic Cu\,K$\alpha_1$ radiation was used for powder
diffraction. Aluminium foil was placed between the sample and
detector to reduce the fluorescent background from iron. The
crystal phases were identified using the JCPDS Powder Diffraction
File. Crystal structures were refined using Rietveld analysis in
the range 20$\degr\!<\!2\theta\!<\!100\degr$ using Rietica
\citep{hunter1998,hill1987}. For angular calibration and
quantitative phase analysis a known mass of high-purity silicon
(9\,\% to 16\,\% by mass depending on the sample) was added as an
internal standard to the powdered sample. Refined structural
variables included lattice parameters, atomic coordinates, metal
site occupancies and isotropic temperature factor; nonstructural
variables included scale factors, background correction and peak
shape. Absorption effects were treated as though they were part of
the overall isotropic temperature factor \citep{scott1981}.

\section{Results}

A total of 24 samples were melted and cooled in the radiation
beam. Backscattered electron microscope images of three samples
are given in Fig.\,\ref{fig:sync}. One sample
(Fig.\,\ref{fig:sync}(a)) has olivine crystals orientated in a
random pattern whereas the sample in Fig.\,\ref{fig:sync}(b) has a
barred olivine texture. The sample in Fig.\,\ref{fig:sync}(c) has
porphyritic microstructure. A compositional X-ray map of a type IA
sample is shown in Fig.\,\ref{fig:samp_xray} and has similar
olivine crystalline structure and composition to the chondrule in
Fig.\,\ref{fig:metxray}.

\begin{figure*}[t]\centering
\includegraphics[width=\textwidth]{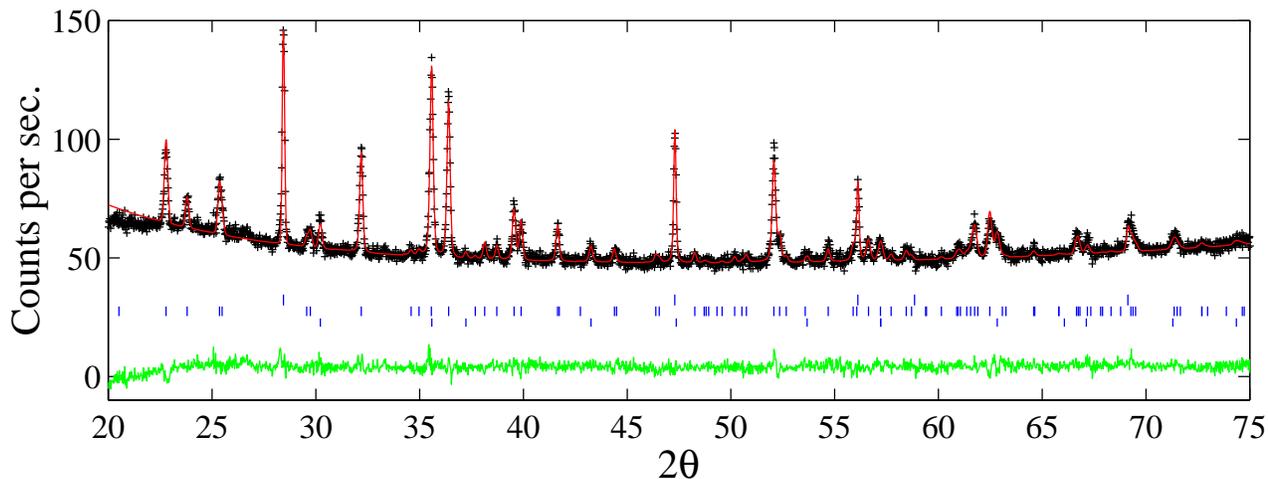}
 \caption{X-ray diffraction pattern of a type II sample. The data
are plotted with crosses and the calculated profile is shown as a
continuous line. The three sets of vertical bars in the middle of
the figure are the calculated reflection positions for silicon
used for calibration purposes (top), olivine (middle) and
magnetite (bottom). The residuals between the data and the
calculated profile are shown at the bottom of the figure
indicating a very good match.}
    \label{fig:reit}
\end{figure*}

There was a general tendency for the microstructure of the samples
to reflect the liquidus temperature \citep{yu1998}. Type IA
samples have the highest liquidus temperature of 1692\,\degr C
(Table\,\ref{tab:comp}). The type IA samples were predominantly
porphyritic in texture with smaller crystals while IAB were about
equal mixture of porphyritic and acicular olivine whereas acicular
olivine predominated in type II. The type IA samples with highest
liquidus temperature had more nuclei throughout the partial melt
at the start of cooling on which the crystals could grow
\citep{rubin2000}. The greater number of nucleation sites resulted
in more crystals with smaller dimensions. Type II samples had the
least number of nucleation sites resulting in fewer but larger
crystals. The faster the cooling rate the more imperfect the
crystals that formed. There is not a perfect match between the
microstructure of chondrules (Fig.\,\ref{fig:met}) and the samples
produced in the synchrotron beam (Fig.\,\ref{fig:sync}). The final
texture of samples depends on a range of factors
\citep{rubin2000,connolly1998,lofgren1990} such as the precursor
composition, maximum temperature, rate of cooling and duration of
heating at maximum temperature. Further experiments are needed to
explore a wider range in parameter space to obtain more
observational constraints on the process.

The Rietveld plot of one type II sample is given in
Fig.\,\ref{fig:reit}. The R values, relating the observations to
the model, for type IA samples were typically R$_\mathrm{p}$ =
6.5\,\% for the profile and R$_\mathrm{wp}$ = 8.2\,\% for the
weighted profile and for type II samples were R$_\mathrm{p}$ =
2.2\,\% and R$_\mathrm{wp}$ = 2.8\,\% indicating very acceptable
fits. The sample in Fig.\,\ref{fig:sync}(a) had olivine with
weight percentage of 56.6\,\% with the remainder being amorphous
while the type II sample shown in Fig.\,\ref{fig:reit} had olivine
(86.2 wt-\%), magnetite (7.2 wt-\%) with the remainder being
amorphous. The Bragg R$_\mathrm{B}$ factors for the individual
crystalline phases were also very acceptable with values less than
4\,\%. Refining the site occupancy factors for the two magnesium
and iron sites allowed the calculation of stoichiometry of the
olivine as Mg$_{1.92}$Fe$_{0.08}$SiO$_{4}$, (i.e. almost pure
fosterite) for the type IA sample (Fig.\,\ref{fig:sync}(a)) and
Mg$_{1.72}$Fe$_{0.28}$SiO$_{4}$ for the type II sample
(Fig.\,\ref{fig:reit}). The higher iron content of the type II
olivine reflects the greater amount of iron in the starting
mixture.

\section{Conclusion}
The experiment demonstrates that GRBs can melt precursor dust
balls to form chondrules in nearby planetary forming systems in
agreement with  predictions. Once formed, the chondrules can move
through the gas more freely and coagulate to form the building
blocks of planets \citep{cuzzi2001}. GRBs with durations greater
than 2\,s are associated with supernovae in massive stars and the
formation of Kerr black holes \citep{popham1999,mcbreen2002}. The
discovery of more than 100 extra-solar giant planets has opened a
range of questions regarding the mechanisms of planetary formation
\citep{santos2003,laughlin2000}. The probability that any
planetary forming system will be blasted by a nearby GRB has been
estimated \citep{mcbreen1999,scalo2002} to be about 0.1\,\%. In
the solar neighbourhood, 7\,\% of stars with high metallicity
harbour a planet whereas less than 1\,\% of stars with solar
metallicity seem to have a planet \citep{santos2003}. The GRB
method seems to be only one of the mechanisms involved because of
the high percentage of stars with planets. There is a further
difficulty for the model in the case of our solar system because
most chondrules were melted more than once requiring a repeating
process \citep{rubin2000,wasson2003}. However the formation of
planetary systems may be enhanced by the presence of a nearby GRB
which will form chondrules across the whole nebula at essentially
the same time \citep{mcbreen1999} but would require a range of
assumptions about the location of the dust in the disk to account
for the properties of chondrules in different classes of
chondrites. Advances in the methods of detecting remnant GRBs and
planetary systems may reveal a link between them in our galaxy.

A GRB can reveal planetary forming systems in other galaxies
because there will be short duration ($\sim$\,1\,hour) bursts of
infrared radiation from the melted dust when chondrules form
across the whole nebula. These infrared bursts can occur for up to
several hundred years after the GRB when the expanding shell of
radiation melts the dust in planetary forming systems that happen
in its way. However the GRB should be in a nearby galaxy to detect
the faint infrared bursts with powerful telescopes such as the
overwhelmingly large telescope (OWL) and the James Webb Space
Telescope (JWST).

\section*{Acknowledgments}

The University College Dublin groups thank Enterprise Ireland for
support. We thank  \AA ke Kvick,  Gavin Vaughan and  Ann Terry for
their help in using the   ID11 beamline at ESRF. Sheila McBreen
and Elaine Winston thank IRCSET for support.

\bibliographystyle{aa}
\bibliography{paper}

\end{document}